\documentclass[useAMS, usenatbib]{mnras}
\usepackage{color}
\usepackage{booktabs}   
\usepackage{tabularx}
\usepackage{float}
\usepackage{totcount}
\usepackage{cuted}
\usepackage[dvipsnames, table]{xcolor}
\usepackage{graphicx}
\usepackage{multirow}
\usepackage[section]{placeins}
\usepackage{subfig}
\usepackage{multicol}
\usepackage{widetext}
\usepackage{upgreek}    
\usepackage{enumitem}   
\setlist[enumerate]{leftmargin=*}   
\setlist[itemize]{leftmargin=*}   
\usepackage[T1]{fontenc}

\usepackage{lipsum} 
\usepackage{amsmath}
\usepackage{amssymb}

\usepackage{txfonts}

\usepackage{xfrac}

\usepackage{tikz,xcolor,hyperref}
\definecolor{lime}{HTML}{A6CE39}
\DeclareRobustCommand{\orcidicon}{
	\begin{tikzpicture}
	\draw[lime, fill=lime] (0,0)
	circle [radius=0.16]
	node[white] {{\fontfamily{qag}\selectfont \tiny ID}};
	\draw[white, fill=white] (-0.0625,0.095)
	circle [radius=0.007];
	\end{tikzpicture}
	\hspace{-2mm}
}

\foreach \x in {A, ..., Z}{\expandafter\xdef\csname orcid\x\endcsname{\noexpand\href{https://orcid.org/\csname orcidauthor\x\endcsname}
			{\noexpand\orcidicon}}
}

\graphicspath{ {figs/} }

\usepackage{xifthen}

\newcommand{\tr}[1]{\textrm{#1}}
\newcommand{\trt}[1]{\textrm{\tiny{#1}}}

\newcommand{\msol}{\tr{M}_{\odot}}

\newcommand{\bigt}{\ensuremath{\uptau}}

\newcommand{\E}[1]{\times\nobreak10^{#1}}

\newcommand{\lr}[2][]{
    \ifthenelse{\equal{#1}{}}{
        {\left(#2\right)}
    }{
        {\left(#2\right)}^{#1}
    }
}

\newcommand{\lrs}[2][]{
    \ifthenelse{\equal{#1}{}}{
        {\left[#2\right]}
    }{
        {\left[#2\right]}^{#1}
    }
}

\newcommand{\scale}[3][]{
    \ifthenelse{\equal{#1}{}}{
        \lr{ \frac{#2}{#3} }
    }{
        {\lr[#1]{ \frac{#2}{#3} }}
    }
}

\newcommand{\sinpar}[2][]{\sin^{#1}\!\lr{#2}}
\newcommand{\cospar}[2][]{\cos^{#1}\!\lr{#2}}

\newcommand{\secref}[1]{\textsection\ref{#1}}
\newcommand{\figref}[1]{Fig.~\ref{#1}}
\newcommand{\refeq}[1]{{Eq.~\ref{#1}}}
\newcommand{\tabref}[1]{{Table~\ref{#1}}}

\newcommand{\torb}{\bigt_\trt{orb}}

\newcommand{\tobs}{T_{\!\trt{obs}}}   

\newcommand{\astropy}{\texttt{Astropy}}
\newcommand{\matplotlib}{\texttt{matplotlib}}
\newcommand{\numpy}{\texttt{NumPy}}
\newcommand{\scipy}{\texttt{SciPy}}
\newcommand{\ipython}{\texttt{ipython}}
\newcommand{\jupyter}{\texttt{jupyter}}

\newcommand{\sympy}{{\texttt{SymPy}}}

\DeclareRobustCommand\kalepy{\texttt{
    {\fontsize{10}{8} \selectfont
    k\kern-0em%
    \raisebox{0ex}{\textcolor{blue}{$a$}}\kern-.14em%
    \textcolor{black}{$l$}\kern-0.05em%
    e\kern+0.04em%
    }%
    p\kern-.05em%
    y%
}}

\newcommand{\pc}{\mathrm{pc}}
\newcommand{\yr}{\mathrm{yr}}        

\newcommand{\fedd}{f_\trt{Edd}}    

\defcitealias{hkm09}{HKM09}
\defcitealias{Begelman1980}{BBR80}
\defcitealias{Rosado1503}{RSG15}
\defcitealias{paper1}{Paper-1}
\defcitealias{paper2}{Paper-2}
\defcitealias{paper3}{Paper-3}
\defcitealias{paper4}{Paper-4}

\def\aj{Astron. J.}              		   		
\def\apj{Astrophys. J.}       		        	 	
\def\apjl{Astrophys. J. Lett.}             		
\def\pasj{PASJ}

\def\pasp{PASP}

\def\apjs{Astrophys. J., Suppl. Ser.}            
\def\mnras{Mon. Not. R. Astron. Soc.}        
\def\prd{Phys. Rev. D}      				
\def\araa{Annu. Rev. Astron. Astrophys.}  
\def\nat{Nature}              				
\def\aap{Astron. Astrophys.}               		

\def\sovast{Soviet Astronomy}
\def\memsai{Mem. Societa Astronomica Italiana}

\newcommand{\per}{p}

\newcommand{\snramp}{S_\tr{\!amp}}

\newcommand{\hbeta}{$H_\beta$}
\newcommand{\fekalpha}{Fe-K$_\alpha$}
\newcommand{\amin}{a_\trt{min}}
\newcommand{\aminsub}[1]{a_\trt{min,#1}}
\newcommand{\amaxdvel}{a_{\trt{max},\Delta \; \!\! v}}
\newcommand{\pmin}{\per_\trt{min}}

\newcommand{\ergps}{\; \textrm{erg s}^{-1}}
\newcommand{\kmps}{\; \textrm{km s}^{-1}}

\newcommand{\lamllamat}[1]{\lambda_\trt{#1} \, L_{\lambda,\trt{#1}}}
\newcommand{\rblr}{R_\trt{BLR}}
\newcommand{\rblrsub}[1]{R_\trt{BLR,#1}}

\newcommand{\rhill}{R_\trt{Hill}}
\newcommand{\rhillsub}[1]{R_\trt{Hill,#1}}
\newcommand{\sigmastar}{\sigma_\star}
\newcommand{\msigma}{$M_\trt{BH}$--$\sigmastar$}

\newcommand{\sigmain}{\sigma_\tr{in}}

\newcommand{\nobs}{N_\tr{obs}}

\newcommand{\nres}{N_\tr{res}}

\newcommand{\sigmablr}{\sigma_\trt{BLR}}
\newcommand{\voffsens}{v_\trt{sens}}
\newcommand{\dvelsens}{{\Delta \; \!\! v}_\trt{sens}}

\newcommand{\dvelminsub}[1]{{\Delta \; \!\! v}_{#1,\trt{min}}}
\newcommand{\dvelmaxsub}[1]{{\Delta \; \!\! v}_{#1,\trt{max}}}
\newcommand{\dvelmin}{{\Delta \; \!\! v}_{2,\trt{min}}}

\newcommand{\dvel}{{\Delta \; \!\! v}}
\newcommand{\voff}{{v_\trt{off}}}
\newcommand{\pvoff}{P_{\voff}}
\newcommand{\pdvel}{P_{\dvel}}
\newcommand{\pvodv}{P_{\voff\dvel}}

\newcommand{\mscale}[1][]{\scale[#1]{M}{10^8 \, \msol}}

\newtotcounter{citnum} 
\def\oldbibitem{} \let\oldbibitem=\bibitem
\def\bibitem{\stepcounter{citnum}\oldbibitem}

\title[Kinematically Offset AGN Binaries]{
    Considerations for the Observability of Kinematically Offset\\Binary AGN
}
\author[L.Z.~Kelley et al.]{
    Luke Zoltan Kelley$^{1}$\thanks{E-mail:lzkelley@northwestern.edu}\orcidA{} \\
    $^{1}$ \begin{minipage}[t]{\linewidth} Center for Interdisciplinary Exploration and Research in Astrophysics (CIERA), and\\Department of Physics \& Astronomy, Northwestern University, Evanston, IL 60208
    \end{minipage}
}

\begin{document}

\maketitle

\begin{abstract}
    The gravitational waves from Massive black-hole (MBH) binaries are expected to be detected by pulsar timing arrays in the next few years.  While they are a promising source for multimessenger observations as binary AGN, few convincing candidates have been identified in electromagnetic surveys.  One approach to identifying candidates has been through spectroscopic surveys searching for offsets or time-dependent offsets of broad emission lines (BLs), which may be characteristic of Doppler shifts from binary orbital motion.  In this study, we predict the parameter space of MBH binaries that should be kinematically detectable.  There is a delicate trade-off between requiring binary separations to be large enough for BL regions to remain attached to one of the AGN, but also small enough such that their orbital velocity is detectable.  We find that kinematic signatures are only observable for the lower-mass secondary AGN, for binaries with total-masses above about $10^8 \, \msol$, and separations between $0.1$ and $1$ pc.  We motivate our usage of a kinematic-offset sensitivity of $10^3 \kmps$, and a sensitivity to changing offsets of $10^2 \kmps$.  With these parameters, and an Eddington ratio of $0.1$, we find that $0.5\%$ of binaries have detectable offsets, and only $0.03\%$ have detectable velocity changes.  Overall, kinematic binary signatures should be expected in fewer than one in $10^4$ AGN.  Better characterizing the intrinsic variability of BLs is crucial to understanding and vetting MBH binary candidates. This requires multi-epoch spectroscopy of large populations of AGN over a variety of timescales.
\end{abstract}

\begin{keywords}
    quasars: supermassive black holes, quasars: emission lines, gravitational waves,  accretion discs
\end{keywords}


\section{Introduction}
    \label{sec:intro}

    Binarity and multiplicity are observed at all astrophysical scales, and binaries of massive black holes (MBHs) are believed to be no exception.  MBHs are almost ubiquitously observed in the centers of massive galaxies \citep{soltan1982, kormendy1995, Magorrian1998} which form through the hierarchical merger of smaller galaxies \citep{Blumenthal1984,Davis+1985}, and continue to merge throughout their lifetimes \citep{Lacey+Cole-1993, Guo+White-0708.1814, Lotz+201108, Newman+2012, rodriguez-gomez2015}.  Two MBHs that are brought together through galaxy merger are only able to form a gravitationally-bound MBH binary (MBHB), and possibly eventually coalesce, through extended dissipative interactions with their local galactic environment \citep[][i.e., dynamical friction, stellar slingshots, circumbinary-disk torques]{Begelman1980}.  Only in a fraction of systems, and over the course of gigayears, do MBHBs reach sufficiently small separations ($\lesssim 10^{-2} \, \pc$) for gravitational-wave (GW) emission to become effective and the system to coalesce \citep{rajagopal1995, sesana2004, milosavljevic2003, merritt2005, paper1}.

    During the final millions of years of inspiral, MBHBs produce GW signals detectable by pulsar timing arrays \citep{sazhin1978, detweiler1979, hellings1983, foster1990} likely within the next few years \citep{Rosado1503, taylor2016, paper2, Mingarelli+2017, paper3}.  A number of electromagnetic signatures of MBHs in binaries have been suggested when one or both MBH is accreting and observable as an active galactic nucleus (AGN) or quasar \citep[e.g.,][]{Komossa2006, Popovic2012, Bogdanovic2015, Kelley+MMA+2019, DeRosa+2019}.  To date, it is still debated whether any `confirmed' binaries have been observed, but the dual radio cores of 4C37.11 \citep{Rodriguez+Taylor+Zavala+2006, Bansal+Taylor+Peck+2017} and the periodic flares of OJ-287 (\citealt{Sillanpaa+Haarala+Valtonen+1988, Valtonen+Lehto+Nilsson+2008, Laine+Dey+Valtonen+2020}; cf.~\citealt{Abraham-2000, Agudo+Marscher+Jorstad+2012, Britzen+Fendt+Witzel+2018}) are increasingly convincing.  A growing sample of AGN in merging galaxies and \textit{dual}-AGN, typically at $\gtrsim$ kpc separations (i.e.~\textit{not} gravitationally bound), have been identified.  These results suggest that AGN activity increases over the course of galaxy mergers \citep{Bennert+2008, koss2012, Comerford201405, Goulding201706}, consistent with galaxy models and observations \citep{Sanders1988, Barnes+Hernquist-1991, barnes1992}.

    Even before the underlying nature of AGN was solidified, it was suggested that the emission lines characteristic of these objects could be used to identify them in binaries \citep{Komberg1968, Gaskell-1983}.  As in a stellar spectroscopic-binary, Doppler shifts due to the orbital motion can create a kinematic offset between the AGN line-center and the host-galaxy rest frame.  Narrow emission lines (NELs), with velocity widths $\lesssim 10^3 \kmps$ and inferred emitting region sizes $\gtrsim 100 \; \pc$ \citep[e.g.,][]{Gaskell-1983b, Antonucci1993}, are likely only produced at distances beyond where a MBH companion would be bound\footnote{Although they are still useful for identifying \textit{dual}-AGN in the same way (e.g.~\citealt{Comerford200810, comerford2012}; cf.~\citealt{Shen201011}).}.  Broad emission lines (BLs) have widths of $\sim 10^4 \kmps$ and emitting regions $\lesssim \pc$ (\textit{Ibid.}), and thus can remain bound and co-move with an AGN inside the binary orbit.  By comparing the centroids of BLs to host stellar absorption or narrow emission lines, many kinematically-offset binary candidates have been put forward \citep{Dotti+200809, Tsalmantza+201106, Eracleous+201106, Decarli+201305, Ju+201306, Shen+201306, Liu+201312, Runnoe+201509}.

    The time variability of BLs has also been used to indicate possible binary motion \citep{Gaskell-1996, Eracleous+201106, Bon+2012, Ju+201306, Shen+201306, Wang+Greene+Ju+2017, Liu+201312, Guo+201809} and to exclude or constrain the presence of companions \citep{Halpern+Filippenko-1988, Eracleous+1997, Liu+201512, Runnoe+201702, Doan+201909, Lu+201910}.  Another possible avenue of detection is systems with double-peaked BLs, which may be produced by the two components of a binary AGN moving relative to each other \citep{Gaskell-1983}.  Some studies have put forward candidates from double-peaked BL sources \citep{Stockton+Farnham-1991, Eracleous+Halpern-1994, Boroson+Lauer-2009, Popovic2012, Decarli+201305, Tsai+2013}, while many others suggest these are more likely produced by single AGN \citep{Eracleous1997, Eracleous+Halpern-2003, Strateva+200307, Storchi-Bergmann+2003, Eracleous+2009, Liu+201512, Doan+201909}.  Keplerian rotation from a disk, for example, is known to produce double-peaks that match observations \citep[e.g.,][]{Eracleous1997}.

    Recently, a number of pioneering theoretical studies have carefully modeled the BL characteristics of binary AGN \citep{Bogdanovic+2008, Shen+Loeb-200912, Nguyen+Bogdanovic-2016, Nguyen+1807.09782, Nguyen+201908}.  Studies using similar BLR-photoionization models have also specifically targeted the reverberation mapping signatures of binary AGN \citep{Wang+2018, Du+2018, Kovacevic+2020}, and changes to flux-ratios of lines due to the presence of a binary \citep{Montuori+2011, Montuori+2012}.  Despite these precision modeling studies, a broader view of the accessible binary population and their parameters has only recently been quantified.  \citet{Ju+201306} and \citet{Pflueger+201803} use models of viscous + GW driven binary evolution to calculate the probability of kinematic detectability based on line-of-sight velocities of each binary component.  In \citet{Pflueger+201803}, the authors find a sweet spot of detectability for mass ratios\footnote{Note that this is likely due in part to their population priors.} $q \equiv M_2/M_1$ between $0.2$ and $0.4$, and binary separations between $10^3$ and $10^4$ gravitational radii, $r_g \equiv G(M_1 + M_2) / c^2$.  Detectable systems constitute $\approx 1\%$ of their overall binary population.

	The goal of the current study is to determine what types of MBHB systems are likely to be detectable from kinematic\footnote{We prefer `kinematic' over `spectroscopic' because non-kinematic, spectroscopic methods of detection exist (e.g.~line-ratios, and spectral deficits from disk gaps).} signatures from optical AGN spectra.  We expand on previous work by 1) considering the competing processes that govern the observability of kinematic offsets, 2) determining the limiting plausible sensitivity of surveys to kinematic signatures, and 3) utilizing self-consistently derived populations of binary AGN.  We emphasize the critical importance of observed BLR sizes, which are not included in the analysis of \citet{Pflueger+201803}.  A lower-limit to viable binary separations is established by requiring that the BLRs remain bound to each AGN.  This criteria leads to an effective maximum velocity offset, which acts against requirements based on survey sensitivity.
	This effect is included in the analysis of \citet{Ju+201306}, but the study does not include a realistic population of systems, and is largely framed in the context of making constraints under the assumption that all AGN are in MBH binaries.

	In our study, we find that the resulting, viable parameter space of detectable binaries is very narrow.  Kinematic signatures are only detectable in the lower-mass secondary AGN, in systems with relatively extreme mass ratios: $q\lesssim 10^{-2}$, and at separations between $0.1$ and $1 \, \pc$ \mbox{($r_g \approx 10^3$--$10^4$)}.  We apply our selection criteria to a population of MBHBs that are derived from cosmological hydrodynamic simulations and evolved using a suite of semi-analytic binary evolution models.  While this population suffers from substantial uncertainties, it is a far more comprehensive and self-consistent approach than has been used for kinematic studies in the past.  From this population we present plausible binary parameters and detection rates that take into account the intrinsic distribution of sources and survey selection effects.  Ultimately, only a very small fraction of all MBHBs are plausibly detectable: one in $\sim 10^{2}$ for velocity offsets, and one in $\sim 10^{4}$ when time-variability is also required.  

\section{Detectable Parameter Space}
    \label{sec:meth_spec}

    \begin{figure}
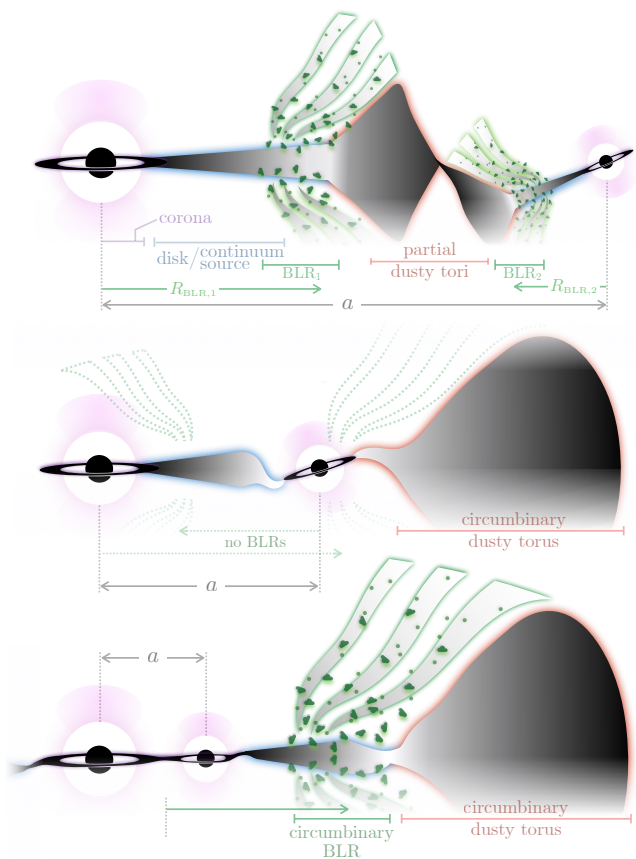

        \centering
        \includegraphics[width=1\columnwidth]{{{figs/comb}}}
        \caption{\textbf{Schematic of Binary AGN and their broad line regions in different regimes.  Top:} $\rblr < \rhill < a$, where BLRs move with each AGN in their orbit.  Here, the dusty tori are near the hill radii and are partially disrupted.  \textbf{Middle:} $\rhill < \rblr \approx a$, where one or both BLRs are disrupted by the binary companion, and the dusty torus is part of the circumbinary disk.  \textbf{Bottom:} $\rblr > \rhill$, both the torus and BLR are in the circumbinary disk and do not follow the orbital motions of the individual AGN.  \textit{Only in the top configuration can offset BLs or changing velocities be detected.}  In either the top or middle panel, the dusty torus may or may not be present.
		Note that the BLR radii (and emission strengths) are proportional to luminosity, which may not be proportional to each AGN's mass.}
        \label{fig:schematic}
    \end{figure}

    \subsection{The BLR Radius and Truncation}
        \label{sec:BLR_radius}

        AGN exhibit fairly tight scaling relationships between their luminosity and the \textit{characteristic}\footnote{Note that the `BLR radius' is weighted by its emissivity, and in the case of reverberation mapping, also its responsivity to continuum-flux variations.  As always, there is also likely bias in the systems with measurements, e.g.~towards high Eddington fractions \citep[e.g.,][]{Brotherton+Runnoe+Shang+2015}.} radius of the BLR \citep{Koratkar+Gaskell-1991,Kaspi+1996,Kaspi+2000}, generally of the form,
        \begin{equation}
            \label{eq:meth_blr_bentz2013}
            \rblr = \alpha \, \scale[\beta]{L}{L_0}.
        \end{equation}
        Relations of this nature are believed to result from the BLR being limited to just within the dusty-torus: determined by the dust sublimation radius, which is proportional to the incident flux \citep[e.g.,][]{Barvainis-1987, Netzer+Laor-1993}.  \citet{Bentz+2013} find best fitting values for the BL–\hbeta{} radius–luminosity relationship as, \mbox{$\alpha_{H_\beta} = 2.8\E{-2} \, \textrm{pc}$} and \mbox{$\beta_{H_\beta} = 0.533$}, where the luminosity is measured at $5100$\AA{} and normalized to \mbox{$L_0 = \lambda \, L_\lambda = 10^{44} \, \textrm{erg s}^{-1}$}.  The radius-luminosity relationship is observed to hold quite well over a broad range of AGN parameters, for example up to eight orders of magnitude in luminosity \citep[e.g.,][]{Kaspi+2007}.  Note that in this section of our analysis, we use best-fitting scaling-relationships without scatter, but the effects of measured scatter are small and discussed later.  Using the bolometric corrections from \citet{Runnoe+2012},
        \begin{align}
            \lamllamat{5100} \approx 2.6\E{45} \, \ergps \left(\frac{M_i}{10^8 \, \msol} \frac{ \, \fedd}{0.1} \right)^{1.1},
        \end{align}
        combined with \refeq{eq:meth_blr_bentz2013} gives,
        \begin{equation}
            \rblr \approx 0.16 \, \textrm{pc} \, \left(\frac{M_i}{10^8 \, \msol} \frac{\fedd}{0.1} \right)^{0.59}.
        \end{equation}

        The BLR must be within the Hill sphere of its MBH such that the emitting portion is not truncated by the companion \citep{Paczynski-1977, Lin+Papaloizou-1979}.  This is shown schematically in \figref{fig:schematic}.  The upper panel shows a wide-separation binary in which the Hill radii of both AGN are outside of the BLR, allowing the BLRs to move with each AGN and thus be observed as kinematically offset.  The bottom panel shows a close-binary configuration in which a single, joint BLR forms outside of the binary orbit.  This `circumbinary' BLR will not track the kinematics of the binary AGN components, and will not produce detectable kinematic offsets or variability.  In between these regimes, at moderate separations, one or neither BLR may be present, which is depicted in the middle panel.

		A fitting formula from \citet{Eggleton-1983} gives the effective Hill radius\footnote{This is the Hill radius defined in a volume-averaged sense.
        How closely the edge of a circumsingle disk approaches the Hill sphere depends on numerous factors, e.g.~viscosity and pressure \citep[e.g.,][]{Paczynski-1977}.} of each object as,
        \begin{equation}
            \label{eq:rhill_eggleton}
            \frac{\rhillsub{i}}{a} = \frac{0.49 q_i^{2/3}}{0.6 q_i^{2/3} + \ln(1+q_i^{1/3})}.
        \end{equation}
        Here, $q_i \equiv \mu_i / (1 - \mu_i)$, is the ratio of object $i$'s mass to that of the other (i.e.~the traditional mass-ratio \mbox{$q = q_2 = M_2/M_1$}), and \mbox{$\mu_i \equiv M_i / M = M_i / (M_1 + M_2)$} is each object's mass-fraction.  This fit is consistent with both more sophisticated analytic and numerical calculations \citep[see, e.g.,][]{Miranda+Lai-2015}.  We use \refeq{eq:rhill_eggleton} for all of our calculations included in the results, but \citet{Artymowicz+Lubow-1994} give a simpler, and quite accurate\footnote{
        The approximation is accurate to better than $95\%$ for $\mu_2 \in [10^{-4}, 0.5]$.  For the primary, the expression is accurate to a factor of two, relative to the \citet{Eggleton-1983} formula, for $1 - \mu_1 \in [10^{-4}, 0.5]$.} relation that is more convenient for analytic calculations:
        \begin{equation}
            \label{eq:rhill_AL94}
            \frac{\rhillsub{2}}{a} = 3^{-2/3} \mu_i^{1/3}.
        \end{equation}
        Using this approximation, we can set a lower-limit on the semi-major axis to preserve the BLR.  Considering that of the secondary MBH in particular,
        \begin{equation}\begin{split}
            \label{eq:amin_blr}
            \aminsub{2} & \equiv 3^{2/3} \rblrsub{2} \, \mu_2^{-1/3} \\
                        & \approx 1.9\E{-1} \, \textrm{pc} \, \scale[0.25]{\mu_2}{0.1} \left(\frac{M}{10^8 \, \msol} \frac{\fedd}{0.1} \right)^{0.59}.
        \end{split}\end{equation}
        We can also express this as a minimum orbital period,
        \begin{equation}
            \pmin \approx 770 \, \textrm{yr} \left( \frac{M}{10^8 \, \msol} \frac{\mu_2}{0.1} \right)^{0.38} \scale[0.88]{\fedd}{0.1}.
        \end{equation}

        \begin{figure}
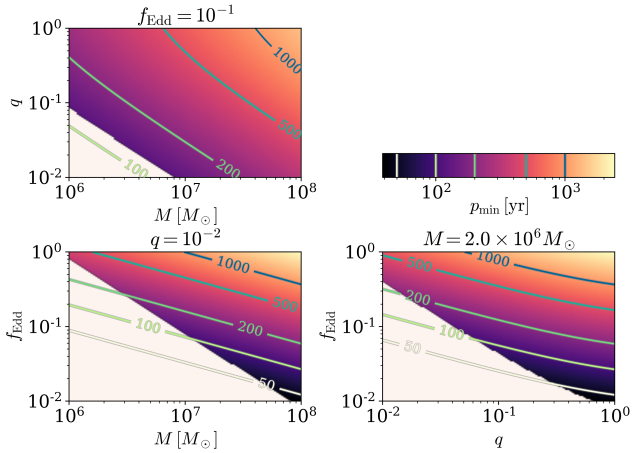

            \centering
            \includegraphics[width=\columnwidth]{{{figs/pmin_sec_lmin+43.00}}}
            \caption{Minimum binary orbital period such that the BLR characteristic radius remains within the Hill sphere of the secondary MBH.  The colormap and contours correspond to orbital periods in years.  The pale regions in the lower-left of each panel are excluded based on a luminosity threshold of $10^{43} \; \textrm{erg s}^{-1}$ (iband $M_{\trt{AB}} \approx 22$).  The fiducial parameters of $\fedd = 10^{-1}$, $q = 10^{-2}$, and $M = 2\E{6} \, \msol$ are chosen to find the smallest orbital periods; such low masses, however, produce very small kinematic offsets that are effectively undetectable.}
            \label{fig:pmin}
        \end{figure}

		The minimum orbital period is plotted in \figref{fig:pmin}, with the additional constraint of a secondary bolometric luminosity above\footnote{We assume a limiting i-band, AB magnitude of $22$ \citep[e.g.,][]{Coffey+2019}, which gives a flux of $\nu_i F_i \approx 2\E{-14} \; \mathrm{ erg/s/cm}^2$.  A redshift of $z = 0.15$, this corresponds to a luminosity of $\approx 10^{43} \; \textrm{erg s}^{-1}$ using the \citet{Runnoe+2012} $5100$\AA{} bolometric correction ($\approx 9.5$).} $10^{43} \; \textrm{erg s}^{-1}$.  Even with optimistic parameters for very low-mass MBHBs, orbital periods under $\sim 100 \, \yr$ are unlikely---and as we discuss later, kinematic offsets from such low masses are likely undetectable.

        The BLR radius used here is a characteristic radius.  BLR photo-ionization models typically find BLRs that span radii, \mbox{$R_\trt{outer} / R_\trt{inner} \approx \rblr / R_\trt{inner} \approx 5$--$10$} \citep[e.g.,][]{Eracleous+1995, Pancoast+201311, Baskin+2014, Baskin+Laor-2018}.  These observations suggest that there is not much room to partially truncate the BLR.  Dust reverberation mapping studies show that the characteristic radius of the BLR is typically a factor of a few ($\sim2$--$5$) within that of the dusty torus \citep[e.g.,][]{Clavel+Wamsteker+Glass-1989, Suganuma+2006, Koshida+2014}.  If emitting BLR-material fills the intervening space, then our usage of the characteristic BLR radius instead of the larger dust-sublimation radius may already be somewhat conservative.  Similarly, numerous studies find that the actual truncation radius of a disk could be $25-50\%$ smaller than the Hill radius \citep[e.g.,][]{Pichardo+2005, Martin+Lubow-2011, Miranda+Lai-2015}.  Eating away at the outer portions of the BLR could substantially diminish the line luminosity.

    \subsection{Observable Offsets}

        A maximum binary separation (minimum velocity-offset) is determined by requiring that the binary components are bound.  This value is determined by the ambient mass density in the galaxy core, which is typically formulated in terms of the nuclear stellar velocity dispersion, $\sigmastar$ --- also a minimum orbital-velocity.  Keeping in mind that the velocity-dispersion measured by a spectroscopic fiber is only a proxy for the dynamically relevant nuclear velocity dispersion\footnote{i.e.~the fiber may cover a larger (or smaller) solid angle of the galaxy than is relevant, and includes projection effects.}, we calculate velocity-dispersion from the \msigma{} of \citet{mcconnell2013},
        \begin{equation}
            \sigmastar = 180 \, \kmps \, \scale[0.177]{M}{10^8 \, \msol}.
        \end{equation}
        This gives a maximum binary separation of,
        \begin{equation}
            \label{eq:amax_bnd}
            a < \frac{GM}{\sigmastar^2} \approx 14 \, \textrm{pc} \,  \scale[0.65]{M}{10^8 \, \msol}.
        \end{equation}
		Again, we note that while scatter in scaling relationships is neglected here, we find that it has little effect our results.  Comparing \refeq{eq:amax_bnd} with \refeq{eq:amin_blr} shows that roughly two dex in separation are viable for kinematic MBHB detection.  Lowering the Eddington ratio can increase this parameter space to include smaller-separation MBHBs, but this comes at the cost of fainter AGN.

        The maximum detectable orbital separation is based on the minimum discernible kinematic velocity-offset.  For now, we consider the optimal case of viewing the system within the orbital plane, and at the phase of peak velocity offset.  Later, we account for varying inclination and orbital phase.  If we parametrize this minimum detectable offset as $\voffsens$, then,
        \begin{equation}\begin{split}
            \label{eq:amax_sens}
            a \lesssim 0.35 \, \textrm{pc} \, \scale{M}{10^8 \, \msol} \scale[-2]{\voffsens}{10^3 \kmps} \scale[2]{1 - \mu_2}{0.9}.
        \end{split}\end{equation}
        Prospects for detecting kinematic offsets are very sensitive to the particular value of $\voffsens$.  A value of \mbox{$\voffsens = 10^3 \kmps$} is a typical selection criterion employed in the literature (e.g.~\citealt{Eracleous+201106}; cf.~\citealt{Liu+201312}), which we explore in \secref{sec:offset_sens}.  Using these fiducial parameters, the maximum separation (\refeq{eq:amax_sens}) is only slightly larger than the minimum separation (\refeq{eq:amin_blr}).  Comparing the two relations suggests a minimum total mass for a detectable binary,
        \begin{equation}
            \label{eq:mass_min}
            \begin{split}
            M_\mathrm{min} \approx 2.3\E{7} \, \msol \, & \scale[4.8]{\voffsens}{10^3 \kmps} \scale[1.4]{\fedd}{0.1} \\
                & \scale[-4.8]{1 - \mu_2}{0.9} \scale[0.61]{\mu_2}{0.1}.
            \end{split}
        \end{equation}
        This expression for the minimum mass scales with almost the 5th power of the velocity-offset sensitivity, an already uncertain parameter, making the usefulness of this relation somewhat shaky.  Note that the strong scaling of the first mass-fraction term makes the minimum mass increase extremely rapidly as the binary approaches equal-mass ($\mu_2 \rightarrow 0.5$).

        \begin{figure*}
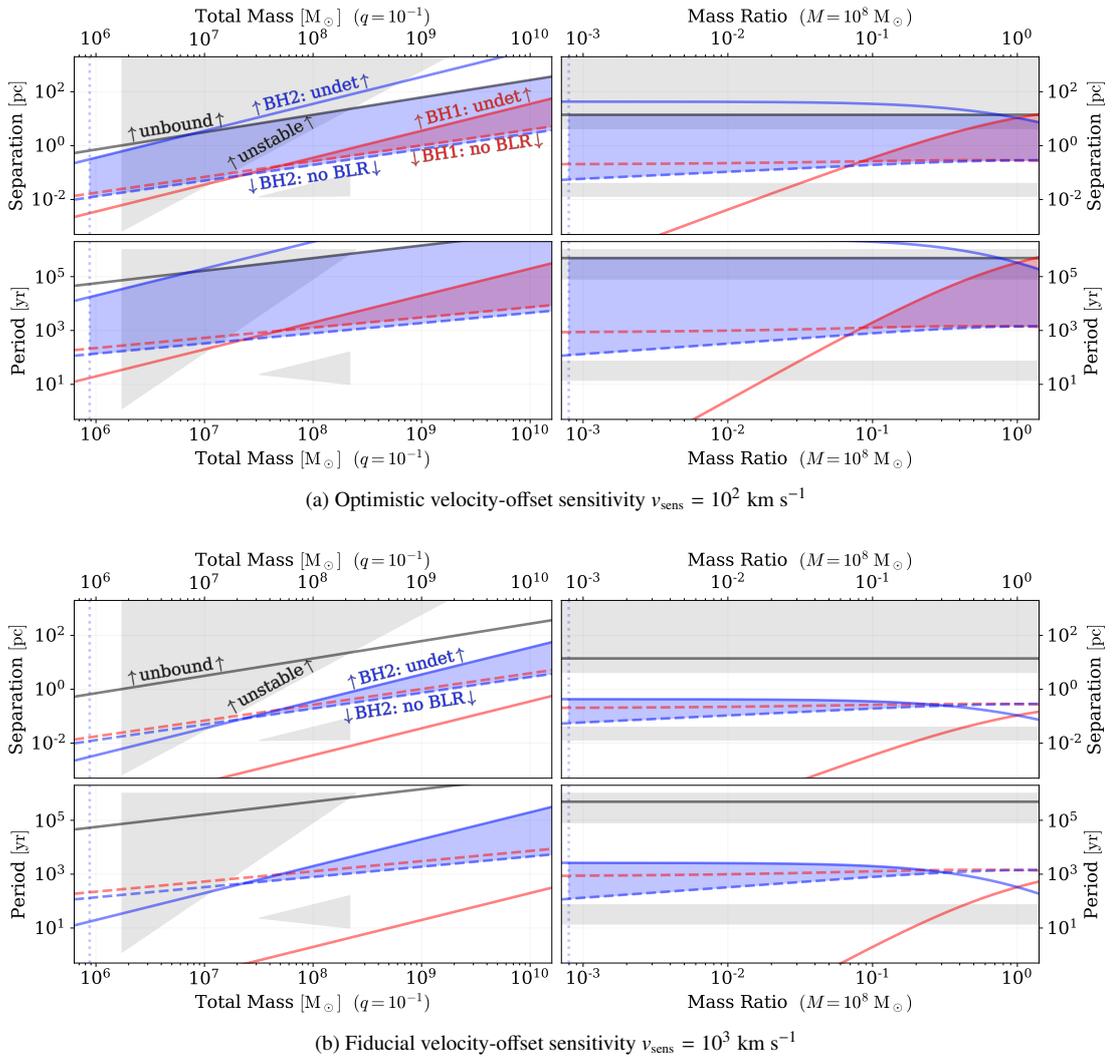

			\centering
            \subfloat[Optimistic velocity-offset sensitivity $\voffsens = 10^2 \kmps$]{{
                \includegraphics[width=1.75\columnwidth]{{{figs/inc_params__mt+8.00_mr-1.00_fedd-1.00_vsens+2.00}}}
            }}
			\\
            \subfloat[Fiducial velocity-offset sensitivity $\voffsens = 10^3 \kmps$]{{
            	\includegraphics[width=1.75\columnwidth]{{{figs/inc_params__mt+8.00_mr-1.00_fedd-1.00_vsens+3.00}}}
            }}
            \caption{\textbf{Binary parameter space accessible by the detection of kinematic BL offsets of the primary (red) and secondary (blue) AGN.}  In each quartet of panels, orbital separation (top) and period (bottom) are plotted against total mass (left; $q=0.1$ fixed) and mass ratio (right; $M = 10^8 \, \msol$ fixed).  We assume an Eddington factor $\fedd = 0.1$ (with a radiative efficiency of $100\%$) and a limiting luminosity of $10^{43} \ergps$.  Regions are excluded based on the criteria derived in \secref{sec:meth_spec}: when the BLR is outside of each component's Hill sphere (`no BLR'; dashed lines), the orbital velocity offset is undetectably small (`undet'; solid lines), the binary is `unbound' (black line), or systems are too faint (vertical, dotted lines).  
			The top set assume a sensitivity to velocity offsets of $\voffsens = 10^2 \kmps$ \citep[e.g.,][]{Liu+201312}, while the bottom set take $\voffsens = 10^3 \kmps$ \citep[e.g.,][]{Eracleous+201106}.  The grey shaded regions are Toomre unstable based on the \textit{single AGN} criteria presented in \citet{hkm09}; how accurately this applies to binaries is unclear.}
            \label{fig:obs_sens}
        \end{figure*}

        The accessible parameter space for detecting kinematic offsets is shown in \figref{fig:obs_sens}.  Each subfigure shows separation (top) and period (bottom) vs. total mass (left) and mass ratio (right).  The shaded regions show where the primary (red) and secondary (blue) kinematic offsets are detectable based on 1) the BLR remaining within the Hill spheres, 2) the binary being bound, and 3) the orbital velocity being larger than $\voffsens$.  The top quartet assume a sensitivity to velocity offsets of $\voffsens = 10^2 \kmps$ \citep[e.g.,][]{Liu+201312}.  The bottom set assume a velocity offset sensitivity of $\voffsens = 10^3 \kmps$ \citep[e.g.,][]{Eracleous+201106}.  Detectable separations are mostly restricted to $\sim [10^{-1}, 10^{+1}] \; \pc$, and orbital periods between $\sim [10^3, 10^5] \; \yr$.  The detectable parameter space is quite insensitive to mass ratio, though this assumes both a fixed accretion rate, and ignores that the secondary AGN may be outshone by the primary (see \secref{sec:disc}).

		There is a subtlety in how the detectable parameter space changes with varying offset sensitivities that is worth emphasizing.  At fixed masses, when $\voffsens$ decreases (improves), the minimum detectable separation ($\amin$) is unchanged (\refeq{eq:amin_blr}), and the maximum detectable separation is increased (\refeq{eq:amax_sens}).  At the same time, however, the minimum detectable \textit{mass} decreases rapidly (\refeq{eq:mass_min}), and thus $\amin$ \textit{over all masses} also decreases.  This can be seen in the upper-left sub-panels of \figref{fig:obs_sens}, where \mbox{$\amin \sim 10^{-1}\,  \pc$} for \mbox{$\voffsens = 10^3 \kmps$}, but decreases to \mbox{$\amin \sim 10^{-2} \, \pc$} for \mbox{$\voffsens = 10^2 \kmps$}.

        Further out in an accretion disk, the velocity gradient decreases until eventually differential rotation is unable to prevent gravitational fragmentation.  Numerous other possible instabilities may also limit the disk extent, and estimations of disk bounds can be highly model dependent.  The presence of a binary adds significant additional complications.  One of the most typically used stability criteria is based on the shear from differential rotation balancing self-gravity---called the Toomre criterion.  We illustrate the possibly-unstable regions as shaded grey in \figref{fig:obs_sens}, based on the calculation from \citet{hkm09},  assuming that the single-AGN Toomre stability criterion is roughly valid for a circumbinary disk.  In their formalism, the disk is decomposed into three, radially-stratified regions distinguished by their dominant pressure source (radiation or thermal) and opacity (Thompson or free-free).  This leads to the two distinct regions of instability in \figref{fig:obs_sens}---triangular when plotted against total-mass, and bands when plotted against mass ratio.

        It is not obvious that Toomre stability near the binary orbit must be required to feed each AGN's circumsingle disk.  However, because we are interested in relatively high-accretion rates $\fedd \gtrsim 10^{-2}$, it seems reasonable to require a stable circumbinary disk.  Requiring Toomre stability at the binary separation is almost always redundant with the sensitivity cut in \refeq{eq:amax_sens}, when using the fiducial $\voffsens = 10^3 \kmps$ (top quartet of \figref{fig:obs_sens}).  This is broadly true over a wide range of total-mass and Eddington-ratios.  When $\voffsens$ is lowered, then smaller binary separations are detectable (as discussed above), but Toomre stability may become an important limitation (bottom quartet of \figref{fig:obs_sens}).  Overall, only systems with total mass $M \gtrsim 10^8 \, \msol$ and separations $a \gtrsim 10^{-1} \, \pc$ are detectable if $\voffsens \sim 10^3 \kmps$, or if the single-AGN Toomre criterion restricts the parameter space.

    \subsection{Observable Changes in Velocity Offsets}
        \label{sec:dvel}

        Consider two spectra taken at times separated by $\tobs \ll \torb$, as will generally be the case; and assume that the observer is oriented within the plane of the MBHB's circular orbit.  We can estimate the time-change of velocity as $\Delta v \approx \lr{dv / dt} \tobs$, where the acceleration $dv/dt$ must be considered at a particular orbital phase.  The maximum change in velocity occurs when both objects are aligned (phase zero for a sinusoidal velocity),
        \begin{equation}
          \label{eq:dvel_max}
          \begin{split}
            \dvelmaxsub{i} & = \tobs \lr[2]{1 - \mu_i} \frac{G M}{a^2}, \\
                \dvelmaxsub{2} & \approx 180 \kmps \, \scale{M}{10^8 \, \msol} \scale{\tobs}{5 \, \yr} \scale[-2]{a}{10^{-1} \, \pc} \scale[2]{1 - \mu_2}{0.9}.
          \end{split}
        \end{equation}
        Detections will be strongly biased against zero phase, however, because it corresponds to zero velocity offset of the projected orbital motion \citep{Eracleous201106}.  When the velocity offset is maximal (phase $\pm \pi/2$), the observed change in velocity is at its minimum:
        \begin{equation}
            \label{eq:dvel_min}
            \begin{split}
              \dvelminsub{i} & = \tobs^2 \lr[3]{1 - \mu_i} \frac{\lr[3/2]{G M}}{a^{7/2}}, \\
                  \dvelminsub{2} & \approx 17 \kmps \, \scale[3/2]{M}{10^8 \, \msol} \scale[2]{\tobs}{5 \, \yr} \scale[-7/2]{a}{10^{-1} \, \pc} \scale[3]{1 - \mu_2}{0.9}.
            \end{split}
        \end{equation}
        While selecting for measurably offset BL centroids, the $\dvelmin$ is a more characteristic value.  Identifying systems with both a substantial velocity offset and velocity shift is significantly more difficult than either alone.

        Consider a survey in which we search solely for BLRs with changing velocities, regardless of whether the BLR centroid is offset or not.  Such a sample would be biased towards the regime in which \refeq{eq:dvel_max} is applicable.  The strong separation dependence initially seems encouraging because even for a modest sensitivity to velocity changes, small separations would still seem to be highly detectable.  Here, however, the separations at which the BLR remains within the Hill spheres of AGN is of critical importance.  By assuming some sensitivity to velocity-changes, $\dvelsens$, we can compare the minimum allowed separation (s.t.~$a > \rhill > \rblr$; Eq.~\ref{eq:amin_blr}) to the maximum separation that is detectable (s.t.~$\dvelmin > \dvelsens$; Eq.~\ref{eq:dvel_min}):
        \begin{equation}
          \label{eq:amax_dv_ratio}
          \begin{split}
            \frac{\amaxdvel}{\amin} \approx 0.70 \, & \scale[1/2]{\tobs}{5 \yr} \scale[-1/2]{\dvelsens}{100 \kmps} \mscale[-0.086] \\
                & \scale{1 - \mu_2}{0.9} \, \scale[-1/4]{\vphantom{1}\mu_2}{0.1} \scale[-1/4]{\fedd}{0.1}.
          \end{split}
        \end{equation}
        For these fiducial parameters, the maximum detectable separation is smaller than the minimum allowed, and there is no range of separations for which a changing velocity-offset is detectable.  Due to the weak scalings with parameters in \refeq{eq:amax_dv_ratio}, this is true for the majority of the plausible parameter space.  Therefore, spectroscopic surveys which target changing BLR velocities are unlikely to detect signatures of binary AGN.  If the binary mass ratio $q\lesssim 10^{-3}$, then a small range of separations with detectable changing-velocities opens up.  Surveys which additionally target offset BLRs are more likely to be in the regime of \refeq{eq:dvel_min}, which makes searches even more challenging.

    \subsection{Double-Peaked BLs}
        \label{sec:doubly_offset}

        For the BLs from both AGN to be observable, the binary orbital velocity must be larger than the combined characteristic widths of both BLs.  As pointed out by, e.g., \citet{Chen+1989, Eracleous+Halpern-1994, Shen+Loeb-2010}, this is generally in conflict with the need for each BLR to remain bound to its MBH and thus be closer-in than the binary orbit itself (see \figref{fig:schematic}).  Kinematically separated BL peaks from each AGN are thus unlikely to ever be observable.  We can estimate the maximum orbital velocity that still preserves each BLR using \refeq{eq:amin_blr},
        \begin{equation}
          v_\trt{orb,max} \approx 1500 \kmps \scale[0.21]{M}{10^8 \, \msol} \scale[-0.13]{\mu_2}{10^{-1}} \scale[-0.29]{\fedd}{10^{-1}},
        \end{equation}
        which is still a factor of two to ten lower than characteristic BL widths, \mbox{$\sigmablr \approx 3\E{3}$ -- $10^4 \kmps$} \citep[e.g.,][]{Stern+Laor-2012}.

        Double-peaked BLs would seem very difficult to be produced from an AGN binary, unless BLs are able to be produced in binary systems from outside of each AGN's Hill radius---while still being bound to each AGN individually.  Allowing the Hill sphere to partially impinge on the BLR modestly increases the maximum allowable orbital velocity, but likely also removes the lower-velocity portion of the BLs, making two distinct peaks more difficult to produce.  A number of insightful reasons why most observed AGN with double-peaked BLs are unlikely to be in binaries are outlined in \citet{Eracleous+Halpern-1994} and \citet{DeRosa+2019}.  Notably, double-peaked BLs are much too common in single AGN \citep[e.g.~20\% of radio-loud AGN;][]{Eracleous+Halpern-2003}, and are well fit by single-AGN models.

    \subsection{The sensitivity to velocity offsets}
        \label{sec:offset_sens}

        The accuracy with which the centroid of a profile can be measured is particularly important in the pulsar timing and stellar/exoplanet communities.  The sensitivity is often calculated using the Cram\'er-Rao bound / Fischer information matrix or equivalently from a maximum likelihood treatment \citep[e.g.,][]{Landman1982, Connes-1985, Butler1996, Lovis+Fischer-2010, Beatty+Gaudi-2015}.  Here we motivate the key scalings based on a simple example.  Consider a triangular spectral feature (and perfectly matching model) of amplitude ($A$) and full width at half maximum (FWHM: $W$).
        Fitting the model to the spectrum with a single pixel (i.e.~resolution element) has a centroid uncertainty $\sigma_i$ due to both the pixel size ($\sigmain$) and the amplitude uncertainty (with root-mean-square amplitude $\sigma_A$).  The amplitude uncertainty translates into a spectral uncertainty based on the feature's slope.  This can be taken as independent and thus adds in quadrature.  Defining the signal-to-noise ratio as $\snramp \equiv A / \sigma_A$, and identifying the number of resolution elements within the feature as $\nres = W / \sigmain$, we can write,
        \begin{equation}
          \begin{split}
            \sigma_i^2 = & \, \sigmain^2 + \lr[2]{\frac{W}{A} \sigma_A} \approx \sigmain^2 \, \scale[2]{\nres}{\snramp}.
          \end{split}
        \end{equation}
        The approximation relies on the feature being wider than it is tall in terms of resolution, or equivalently that the SNR per pixel is small.  We can convert from FWHM to Gaussian standard-deviation, and identify this with the characteristic width of the BL such that, $W = 2 \lr[1/2]{2 \ln 2} \sigmablr$.  The combined accuracy of fitting with $\nres$ elements and $\nobs$ independent observations can then be written as,
        \begin{equation}
          \label{eq:acc_simple}
          \begin{split}
            \sigma_v & = \lr[1/4]{2 \ln 2} \frac{\lr[1/2]{\sigmablr \sigmain} \!\!\!\!}{S} \\
                & \approx 10^2 \, \kmps \, \nobs^{-1/2} \scale[-1]{\snramp}{10^{1/2}} \scale[1/2]{\sigmablr}{10^4 \kmps} \scale[1/2]{\sigmain}{10 \kmps}.
          \end{split}
        \end{equation}
        The derivation leading to \refeq{eq:acc_simple} is simple and highly idealized, but in most respects it should represent an \textit{optimal} sensitivity.  BLs are of course not triangles, and in general are centrally-peaked---with the highest amplitude feature being much narrower, and contributing most of the SNR.  The highest velocity material, on the other hand, which should best trace the offset AGN components, is much broader and contributes less to the SNR, which make the effective accuracy much lower.  BL shapes are also known to be typically asymmetric and highly variable, both of which further degrade the accuracy of centroiding.

        The best sensitivity is likely achieved by template matching, using a template constructed from independent observations to decrease the impact of uncorrelated temporal variations.  \citet[][Eq.A3]{Cordes+Shannon-2010}, for example, give the minimum centroid error of template matching for pulsar pulses and find an expression matching the scalings of \refeq{eq:acc_simple}, and with a numerical pre-factor that differs by only $\approx 5\%$.  Their result is also identical to the traditional maximum-likelihood based estimates, for example in \citet[][Eq.~4b]{Landman1982}, and should also be effectively very similar to the results of cross-correlation which are typically used for BLs. \citet{Beatty+Gaudi-2015} show that variations to profile shape (e.g.~Gaussian vs.~Lorentzian, etc) lead to changes in the prefactor on the order of $\sim 10\%$.  Overall, this suggests that there is a fairly clear theoretical limit to centroid accuracy that is quite general and fairly insensitive to assumptions.  

        Here we have assumed an intrinsic BL velocity-width, $\sigmablr \approx 10^4 \kmps$.  A somewhat lower value of $\approx 3\E{3} \kmps$ is more representative of the bulk of observed AGN \citep[e.g][]{Stern+Laor-2012}.  At the same time, the tails of BLs beyond the FWHM are produced closer to the AGN, and thus likely the best tracers of its motion.  `Instrumental broadening' in the literature is also typically much larger \citep[e.g.~$\gtrsim 100 \kmps$][]{Barth+2015} than our fiducial value of $\sigmain = 10 \kmps$, though determining and comparing the most appropriate metric (i.e.~systemic vs.~per-pixel, at what SNR, etc) is not entirely obvious, and of course varies by instrument and methodology.


        The preceding error analysis suggests that the accuracy to which AGN BL positions can be measured is $\sim 100 \kmps$, at best.  BLRs, however, are highly dynamical, and the ability to identify kinematic offsets as potential binary candidates also depends on the intrinsic stability of BL signals themselves.  Velocity offsets of broad Balmer lines tend to vary (`jitter') over a few hundred $\kmps{}$ over time intervals of $1-6$ months \citep{Barth+2015, Doan+2019}.  Jitter amplitudes from \citet{Doan+2019} are shown in \figref{fig:jitter_doan2019}, where the median jitter for red and blue components are $270$ km/s and $390$ km/s respectively.


        A possible source of this jitter is that varying continuum emission reverberates off of BLR components whose Doppler factors vary asymmetrically with their time-delay \citep{Blandford+McKee-1982}.  For example, consider an AGN whose BLR is part of a fast outflow, oriented in a disk-like geometry that is inclined relative to the observer.  If the continuum emission increases in brightness, the observer will first see it reverberate off of the nearer-side of the disk, which is blue-shifted from the outflow, and thus the early response will be a bluing of the line, followed later by a reddening.  \citet{Barth+2015} discuss numerous possible sources of jitter, including this ``asymmetric reverberation'' induced jitter, and find that it can produce variations of hundreds of $\kmps$ on timescales comparable to the continuum variability.

        AGN in MBH binary systems are expected to be intrinsically rare, with a fraction of all AGN at best $\lesssim 10^{-2}$ \citep[e.g.,][]{Volonteri2009, Kelley+2019}.  Naively, a characteristic kinematic offset should be at least this rare ($10^{-2} \approx 2.3$ Gaussian standard-deviations) in single AGN to hint at the presence of a binary.  This implies a minimum velocity offset of $\gtrsim 230 \kmps$ from \refeq{eq:acc_simple}.  The observational jitter measurements of \citet{Doan+2019} imply a significantly higher empirical cutoff of \mbox{$\gtrsim 600$--$800 \kmps$}, and the true high-jitter tails of the distribution are likely significantly larger than based on these $\approx$ dozen measurements.

        Based on these considerations, taking \mbox{$\voffsens \sim 10^3 \kmps$} as a typical intrinsic sensitivity to kinematic offsets seems very reasonable, and is often used in the literature to select for offset systems \citep[e.g.,][]{Eracleous+201106}.  Also, using this value for the sensitivity to changing velocity offsets ($\dvelsens$) would seem similarly reasonable, whereas \mbox{$\dvelsens \sim 10^2 \kmps$} would seem quite optimistic except, perhaps, in cases of very high SNR spectra, and AGN known to have BLs that are particularly stable.  The latter, however, might itself substantially bias against binary AGN.  The presence of a massive companion will generally seed and grow time-dependent variability in the accretion flow.  While this is likely concentrated at timescales near the orbital period \citep[e.g.,][]{Miranda+Munoz+Lai-2017}, faster variations are certainly possible \citep[e.g.,][]{farris201310}, for example if spiral density waves are excited in each AGN's circumsingle disk.


\section{Kinematic Offsets in Evolving MBH Binary Populations}
    \label{sec:res}

    \begin{figure*}
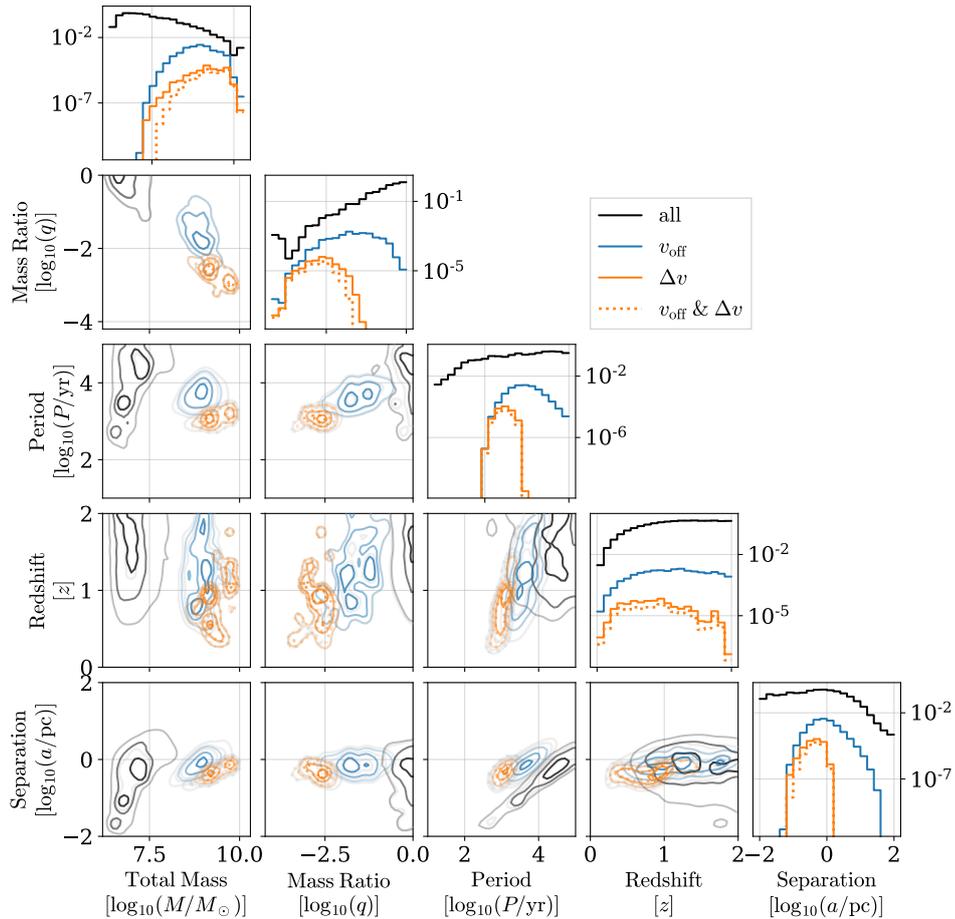

        \centering
        \includegraphics[width=1.5\columnwidth]{{{figs/det-nums_fedd-1.00_voff+3.00_dvel+2.00_tobs+0.70}}}
        \caption{\textbf{Distributions of parameters for all binaries (black) and those with a detectable kinematic signature in the secondary AGN (colors).}  Less than $10^{-7}$ of primaries are observable, which are not shown.  Secondary AGN with observable velocity offsets ($\voff$) are shown in blue assuming a sensitivity of \mbox{$\voffsens = 10^3 \kmps$}.  Secondaries with observable changing velocities ($\dvel$) are shown in orange (solid) for a sensitivity $\dvelsens = 10^2 \kmps$, and observing baseline of \mbox{$\tobs = 5\, \yr$}.  Systems where both offsets and changes are detectable (orange, dotted) are mostly a uniform subset of the $\dvel$ sample, and are almost indistinguishable in the 2D contour plots.  An optical flux cutoff of \mbox{$6\E{-14} \mathrm{erg/s/cm}^2$} (AB Mag $\approx 21$) is also imposed, where all AGN are assumed to be accreting at an Eddington fraction of $0.1$.}
        \label{fig:det_bins_num}
    \end{figure*}

    \begin{figure}
        \centering
        \includegraphics[width=1\columnwidth]{{{figs/det-fracs_3pars_fedd-1.00_voff+3.00_dvel+2.00_tobs+0.70_zoom}}}
        \caption{\textbf{Fraction of binary AGN with detectable kinematic signatures from the secondary.}  The same parameters are used as in \figref{fig:det_bins_num}: \mbox{$\voffsens = 10^3 \kmps$}, observing baseline \mbox{$\tobs = 5\, \yr$}, $\dvelsens = 10^2 \kmps$, and \mbox{$F_\trt{opt} > 6\E{-14} \mathrm{erg/s/cm}^2$} ($M_\trt{AB} \gtrsim 21$).  Overall, $0.49\%$ of secondaries have a detectable offset and only $0.027\%$ have detectable changes in velocity.  Both kinematic signatures are simultaneously detectable in $0.014\%$ of systems.  Distributions of binary parameters for detectable systems are listed in \tabref{tab:obs_pars}.}
        \label{fig:det_bins_frac}
    \end{figure}

    The binary parameter space that is accessible through kinematic offsets was shown in \figref{fig:obs_sens} and discussed above.  The intrinsic distribution of binary parameters is biased towards low total masses and large separations \citep[see, e.g.,][]{paper1}.  To account for the underlying parameter distributions, we use a simulated population of binary AGN and apply the selection criteria described in \secref{sec:meth_spec}.  The binary population is derived from the `Illustris' cosmological, hydrodynamic simulations \citep{vogelsberger2013, vogelsberger2014a, torrey2014, vogelsberger2014b, genel2014, sijacki2015}, and evolved in post-processing using semi-analytic models presented in \citep{paper1, paper2}.  This produces a population of $\approx 10^4$ binaries, where each component mass is between $10^6$ and $\approx 10^{10} \; \msol$.

    Based on the cosmological evolution of the universe, we can extrapolate from this population in a finite volume
	to an entire population of binaries in the observer's past light-cone.  We employ kernel density-estimation (KDE), a type of non-parametric multi-dimensional parameter estimation, to infer the underlying distribution of binary properties, and then resample to the required tens to hundreds of millions of systems.  The KDE approach \citep[e.g.,][]{Scott+Sain-2005} smooths each data point over a `kernel' that fills in gaps between the data points, while still preserving the covariance and structure of the input.  Resampling from a histogram is equivalent to a KDE with a top-hat kernel, at discrete locations.  The KDE package we use is being generalized for use by the community \citeauthor{kalepy2020} (\textit{in prep.}).

    The accretion rate onto each component of the binaries is not resolved in Illustris (nor the structure of their disks or BLRs), although it does provide an accretion rate onto the combined system.  How this overall accretion rate into post-merger galactic nuclei is partitioned between binary MBHs is unclear, particularly for the binary separations of interest here (see \secref{sec:disc}).  Due to these uncertainties we continue to adopt a fixed Eddington factor 
	To account for viewing angle effects for kinematic offsets ($\voff$) and their shifts ($\dvel$), we use the expressions derived in \citet{Pflueger+201803} for the probability of detecting a given offset $\pvoff(x \equiv \voffsens / \voff)$ or shift $\pdvel(y\equiv \dvelsens / \dvel)$,
    \begin{equation}
        \begin{split}
            \pvoff = 1 \, - \, & \frac{2}{\pi} \lrs{\arcsin x + x \ln \lr{\frac{1 + \cospar{\arcsin x}}{x} } \,}, \\
            \pdvel = \, & \frac{2}{\pi} \lrs{\arccos y - y \ln \lr{\frac{1 + \sinpar{\arccos y}}{y}} \,}.
        \end{split}
    \end{equation}
    Because $\pvoff$ and $\pdvel$ are strongly anti-correlated in viewing angle (\secref{sec:dvel}), we use the approximation that the combined probability \mbox{$\pvodv = \pvoff \pdvel$}, which gives accurate results\footnote{Note that \citet{Pflueger+201803} do provide an exact expression for $\pvodv$, though it must be integrated numerically for each parameter combination.  The approximation we use is typically accurate to $\sim 10\%$ for individual systems, and better than $5\%$ for the overall population.}.

    The distribution of binary parameters are shown in \figref{fig:det_bins_num} for all binaries and those with detectable kinematic signatures.  The populations are compared for all binaries (black), those with a detectable kinematic offset ($\voff$, blue), changing offset ($\dvel$, orange solid), and both (orange dashed).  Because, generally, $\pdvel \ll \pvoff$, half ($52\%$) of systems with a detectable changing-offset also have offsets $\voff > \voffsens$.  All binaries, and those with detectable $\voff$, and $\dvel$, are each concentrated in different regions of parameters space.  All binaries tend to have low total masses, high-mass ratios\footnote{Note that this is partially a selection effect because we only include MBH masses $M > 10^6 \, \msol$, and thus the dominant number of binaries with total-masses near $M \sim 10^6 \, \msol$ can only have near-equal mass ratios.}, and long orbital periods where their residence times $a / (da/dt)$ are longest.  The decline in systems with $a \gtrsim 10 \, \pc$ is due to systems not being gravitationally bound at larger separations.  All detectable systems are biased towards the largest total masses, and more extreme mass-ratios, which produce the largest orbital velocities in the secondary AGN.  For the same reason, detectable systems prefer smaller separations --- but below $\sim 0.1 \, \pc$, the shrinking Hill radii begin to exclude BLRs.  Only one in roughly every four million binaries has a detectable primary offset, and these never have a detectable changing offset.  None of our binaries have double-peaked BLs with offsets in both the primary and secondary AGN.  We therefore focus our analysis on systems with a detectable secondary.

    The fraction of binaries that are kinematically detectable are plotted in \figref{fig:det_bins_frac} for a subset of parameters.  Overall, $5\E{-3}$ of binaries have a secondary AGN with detectable $\voff > 10^3 \kmps$, and only $3\E{-4}$ with $\dvel > 10^2 \kmps$.  Parametrically, systems with detectable $\dvel$ are a fairly uniform subset of those with detectable $\voff$.  The $\dvel$ and $\voff$ are necessarily anti-correlated in orbital phase (\secref{sec:dvel}).  For this reason only half of $\dvel$ detectable systems also have a detectable $\voff$, at $1.4\E{-4}$ of all binaries.  We have also tried including the scatter from observational scaling relationships in our calculations, in particular the scatter in \msigma{} from \citet{mcconnell2013} and BL–\hbeta{} radius–luminosity from \citet{Bentz+2013}.  The changes to the plotted histograms are imperceptible, and the overall detection rates are hardly changed: systems with detectable $\dvel$ go from $2.7\E{-4}$ to $2.8\E{-4}$, and those with both $\dvel$ and $\voff$ go from $1.4\E{-4}$ to $1.5\E{-4}$.

    Our binary evolution models take into account the mass ratios, both of the MBHs and their host galaxies, in all stages of their evolution.  Still, the dynamics of more extreme mass-ratio systems are even more uncertain than that of MBHBs overall.  Our analysis also does not take into account that the secondary AGN, which always produced the detectable kinematic offsets, must be discernible behind the emission of the primary, which may be much brighter.  For these reasons we caution that the number of detectable extreme mass-ratio systems may be overestimated here.

    As apparent in \figref{fig:det_bins_num}, detectable binary parameters are only very weakly dependent on redshift.  The H$\beta$ BL is one of the most commonly used to search for binary candidates, but is only observable in SDSS out to a redshift of $\approx 0.7$;, though MgII and CIV are useful at larger redshifts.  In \tabref{tab:obs_pars} we include tabulated quantiles for detectable systems at low to moderate redshifts: $z < 0.7$.

\section{Discussion \& Conclusions}
    \label{sec:disc}

    Based on empirical scaling relations in AGN BL properties, and simple physical considerations, we have shown that kinematically offset AGN are detectable in only a narrow range of parameter space.  BL regions from relatively bright AGN are observed to have characteristic radii on the order of $0.1 \, \pc$ \citep[e.g.,][]{Kaspi+2007, Bentz+2013}, suggesting that binary separations must be at least this wide for their BLs to remain observable, and to trace the motion of each AGN component.  At binary separations larger than $\sim 1 \pc$, orbital velocities become too small to reliably detect.  For the same reason, binaries of total mass much less than $\approx 10^8 \, \msol$ are rarely detectable.  Even if we have underestimated the sensitivity of spectroscopic surveys to kinematic offsets, the accretion disk radii corresponding to BLRs in lower mass systems may be susceptible to gravitational instability.  The dynamical stability of disks, especially in the case of circumbinary disks, however, is highly uncertain.

    Requiring a minimum binary separation implies that the shortest orbital periods of kinematically detectable systems is optimistically hundreds of $yr$, with periods of $10^3$--$10^4 \, \yr$ being typical.  While binary candidates from AGN with photometric, periodically-variability can be filtered to some degree by waiting for numerous complete cycles \citep[e.g.,][]{sesana201703}, likely this cannot be done for kinematic candidates.  

    Much of our analysis hinges on the premise that BLs are only observable if produced at the measured, characteristic radii of BLRs.  While these measurements span a wide range of AGN parameters \citep[e.g.~redshift and luminosity;][]{Kaspi+2007}, the total number of systems by which they are calibrated is still relatively small and may suffer from selection biases.  How BLs are formed in more complex environments, such as MBH binaries and their circumbinary- and circumsingle- disks are only beginning to be explored in detail \citep[e.g.,][]{Bogdanovic+2008, Shen+Loeb-200912, Nguyen+Bogdanovic-2016, Nguyen+1807.09782}.  Studies such as these are starting to show how intricate the BLR environment can be around binary systems.  In our analysis, and all others on binary AGN BLs to our knowledge, feedback processes such as energetic outflows and jets have been entirely neglected.  These processes could not only alter but even disrupt a companion's BLR (either dynamically or radiatively).

	The same considerations apply to dusty tori immediately outside of BLRs.  The absence of dusty tori may be characteristic for AGN with binary companions (see middle-panel of \figref{fig:schematic}), though this is highly speculative.  If it is the case, it may be detectable through an infrared deficit.  Obscuration of the BLR for viewing angles near the plane of the AGN disk, one of the hallmarks of AGN unification (e.g.,~\citealt{Antonucci1985, Antonucci1993, Urry1995, Urry-2004, Nenkova+2008}; but, cf.~\citealt{Penston+Perez-1984, DiPompeo+2017}), has also been neglected in our analysis.  Dusty torus truncation may remediate some obscuration effects, but not if they are instead produced by high column-densities from dusty outflows \citep[e.g.,][]{Konigl+Kartje-1994, Elitzur+Shlosman-2006}, or time-dependent but more isotropic obscurers.  Galaxy mergers also trigger increases in AGN obscuration \citep[e.g.,][]{Hopkins+Hernquist+Martini-2005, Koss+Assef+Balokovic+2016, Ricci+Bauer+Treister+2017, Blecha+Snyder+Satyapal+2018, Pfeifle+2019}, though how the timing and duration of increased obscuration compares to MBHB inspiral times and the lifetimes of kinematically-detectable AGN is unclear.

    Our analysis is sensitive to the fundamental accuracy by which BL centroids can be measured.  Based on highly-simplified profile fitting considerations, we find that accuracies are likely at best $\sim 100 \kmps$.  BLs are known to be asymmetric and highly variable, where studies have shown that their velocities can shift by many hundreds of $\kmps$ on month to year timescales \citep{Barth+2015, Doan+2019}.  Based on these considerations, we use a fiducial sensitivity to velocity offsets of $\voffsens = 10^3 \kmps$, and what we consider to be a very optimistic sensitivity to changing velocities of $\dvelsens = 10^2 \kmps$.  These considerations do not account for possible contamination of offset lines produced by post-merger, recoiling AGN \citep[e.g.,][]{blecha2016, Sayeb+2020}; but ultimately, in that case, it would still mark the presence of a \textit{past} binary.  The amplitude of velocity shifts from Brownian motion of single-AGN are expected to be very small \citep[e.g.,][]{Merritt-2001}, at least in relatively massive galaxies\footnote{though wandering and spatially-offset MBH are a rapidly developing field of research, e.g., \citet{Tremmel+Governato+Volonteri-2018} and \citet[][cf.~\citealt{Eftekhari+Berger+Margalit+2020}]{Reines+Condon+Darling+2020}}, and thus should not be an important confusion or noise source.

    We apply our observability criteria to a population of Illustris MBH binaries with masses between $\approx10^6$ -- $10^{10} \, \msol$, that are evolved from galaxy merger at kpc-scale separations down to eventual coalescence \citep{paper1, paper2}.  Using the fiducial sensitivities discussed above, and assuming a universal Eddington fraction of $0.1$, we find that the secondary AGN should have a detectable kinematic offset in $0.5\%$ of binaries.  Changes in the BL velocity are only detectable in $0.03\%$ of systems, with roughly half of those also having a detectable offset from the rest frame.  Detectable systems have binary separations near and just below $1 \, \pc$, total masses $\gtrsim 10^9 \, \msol$, and extreme mass-ratios, typically with $q \lesssim 10^{-2}$.  We caution that the evolution of these extreme mass-ratio systems is especially uncertain.  For example, if accretion strongly favors the secondary, then mass ratios may evolve noticeably during inspiral \citep{Siwek+Kelley+Hernquist-2020} which may decrease the occurrence rate of binary AGN with $q \lesssim 0.1$.  Given our model assumptions, and the overall fraction of AGN in binaries being \mbox{$10^{-2}$ -- $10^{-3}$} \citep{Volonteri2009, Kelley+2019}, less than one in $10^4$ AGN should have detectable kinematic signatures from binary motion.  These results are not significantly effected by the scatter in observed scaling relationships.

	One of the key differences between this analysis and previous ones is the consideration of the characteristic radii of optical BLRs: if the observed scalings hold in binaries, then the binary separation must be large enough that the BLR remains within the Hill sphere of each component AGN.  This requirement is in tension with the need for small orbital separations to produce detectably large velocity shifts.  The parameters used in this study are motivated primarily by studies of the \hbeta{} line.  Other optical BLs (e.g.~MgII, CIV), show very similar characteristics \citep[e.g.,][]{Homayouni+Trump+Grier-2020}, and the same overall conclusions likely apply.  X-ray lines, like \fekalpha{}, which are produced in the immediate vicinities of MBHs, can have significantly larger Doppler shifts, are unaffected by truncation from the companion MBH, and may tend to be brighter in the secondary AGN.  Especially in light of proposed X-ray missions such as Athena \citep{Athena2017}, Lynx \citep{Lynx2018}, and XRISM \citep{XRISM2020}, prospects for detecting MBH binaries through X-Ray observations \citep[e.g.,][]{Yu+Lu-2001, McKernan+2013, McKernan+Ford-2015, Haiman-2017} should continue to be explored.

    We find that the primary AGN in binaries are effectively undetectable.  We do not account for whether or not the secondary AGN, which are significantly lower in mass than the primaries in detectable systems, are able to produce BLs bright enough to outshine the primary.  Many studies have found that in circumbinary systems, accretion preferentially favors the secondary \citep{Artymowicz+Lubow-1994, gould+rix-2000, farris201310}.  As pointed out in \citet{Pflueger+201803}, this does not necessarily mean the secondary will also have a brighter BL, as the Hill sphere (and thus secondary accretion disk and BLR) is smaller than the primary.  Indeed, the analysis in \citet{Nguyen+201908, Nguyen+1807.09782} suggests that even if the secondary is accreting preferentially, the primary dominates the \hbeta{} line, making detection of kinematic signatures in the secondary far more difficult.

    A number of reasons why double-peaked BLs are unlikely to be produced by the two BLs of a binary AGN are outlined by Eracleous et al. (and others), particularly in \citet{Eracleous+Halpern-1994}.  Ultimately, it is very difficult to construct binary parameters in which the difference in orbital velocities between components can reach the width of the primary's BL.  We find no systems in our population where this occurs.  Theoretical considerations of disk alignment suggest that circumsingle disks (and thus possibly BLRs) may tend to be coplanar with the binary orbit \citep[e.g.,][]{Artymowicz+Lubow-1994, Ivanov+1999, bogdanovic2007, Dotti+2010, Miller+Krolik-2013}.  If this is the case, and if a dusty torus is still present in the system, obscuration could be an even worse issue.  If both components of the binary contribute noticeably to a combined BL, then the primary may not only mask the brightness of the secondary's BL, but its opposing velocity may negate some of the apparent velocity shift.

    Unfortunately, our analysis has yielded a fairly pessimistic forecast for the detectability of binary MBHs through kinematic offsets in AGN optical spectra.  Our ability to identify binary AGN in this way rests crucially on being able to better characterize the intrinsic asymmetry and variability of BLs and their production regions.  In addition to existing data from reverberation mapping campaigns, a larger sample of AGN with multi-epoch spectroscopy, over a variety of timescales, is needed to understand what produces the `jitter' in AGN BLRs.  Those variations, especially if they are produced in response to continuum fluctuations, may be able to be characterized and modeled-out when searching for binary signatures.  A campaign like the BH-Mapper \citep[part of SDSS-V,][]{Kollmeier+201711} could provide invaluable constraints, and significantly increase the population of AGN required to plausibly identify the signposts of binarity which appear to be so rare.

\section*{Acknowledgments}
	I am very thankful to Claude-Andr\'e Fauche-Giguer\'e, Diego Mu\~{n}oz and Jonathan Stern for frequent consultations and advise.  Laura Blecha, Tamara Bogdanovi\'c, Claude-Andr\'e Fauche-Giguer\'e, Jessie Runnoe, and Jonathan Stern provided thorough and invaluable feedback on an early draft.  I also thank Maria Charisi and Joseph Simon for encouraging and stimulating conversations on the subject of kinematic binary detections.

    This research made use of \astropy, a community-developed core Python package for Astronomy \citep{astropy2013}, in addition to \scipy~\citep{scipy}, \ipython~\citep{ipython}, \jupyter~notebook~\citep{jupyter}, \numpy~\citep{numpy2011} \& \sympy~\citep{sympy2017}.  All figures were generated using \matplotlib~\citep{matplotlib2007}.  Kernel density estimation was performed using the \kalepy{} package (\href{https://github.com/lzkelley/kalepy}{github.com/lzkelley/kalepy}) \citep{kalepy2020}.

    The Illustris data is available online at \href{https://www.illustris-project.org/}{www.illustris-project.org} \citep{nelson2015}, and Illustris-TNG data at \href{https://www.tng-project.org/}{www.tng-project.org} \citep{Nelson+1812.05609}.

\let\oldUrl\url
\renewcommand{\url}[1]{\href{#1}{Link}}

\quad{}
\bibliographystyle{mnras}
\bibliography{references,refs_agn}

\onecolumn
\clearpage

\appendix

    \section{Additional Material}
        \label{sec:app_figs}

        \noindent\begin{minipage}{\linewidth}

            \centering
            \includegraphics[width=0.55\columnwidth]{{{figs/doan+2019_jitter}}}
            
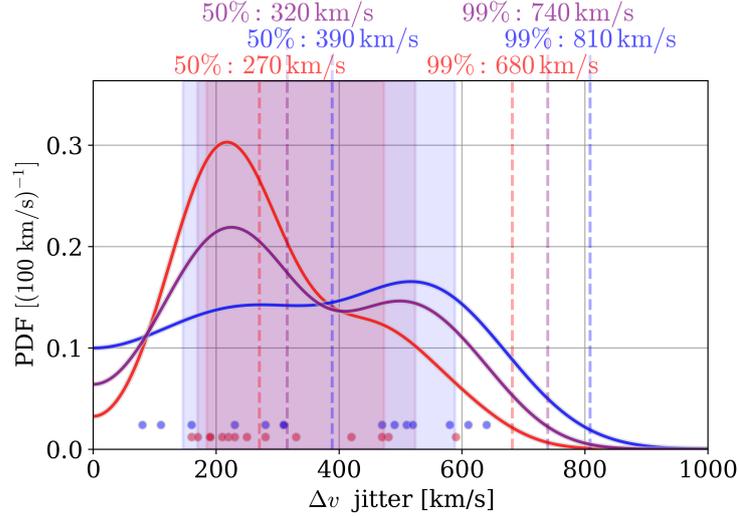
\captionof{figure}{Distribution of velocity-offset variations (`jitters') from \citep{Doan+2019}, for the red and blue components of the observed BLs.  Measurements are shown as ticks at $y=0$, and KDE distributions are shown using a Scott's factor bandwidth, and Gaussian kernel.  The purple curve is the KDE distribution taking both red and blue data as independent.  The KDE $50\%$ (median) and $99\%$ (Gaussian standard deviation $\sigma \approx 2.3$) jitter for each component are shown with dashed vertical lines and corresponding labels.}
            \label{fig:jitter_doan2019}

            \vspace{10ex}

            \setlength{\tabcolsep}{8pt}
            \begin{tabular}{p{1.5cm} c c c c c c c}
                 & & & $5\%$ & $25\%$ & $50\%$ & $75\%$ & $95\%$ \\ [0.5ex]
                \hline
                \multirow{4}{*}{All}  & \multirow{4}{*}{-}
                    & $M \; [M_\odot]$ & $3.4 \times10^{ 6 }$ & $7.2 \times10^{ 6 }$ & $1.8 \times10^{ 7 }$ & $6.0 \times10^{ 7 }$ & $4.1 \times10^{ 8 }$ \\
                    & & q & $2.9 \times10^{ -2 }$ & $1.7 \times10^{ -1 }$ & $4.0 \times10^{ -1 }$ & $6.8 \times10^{ -1 }$ & $9.3 \times10^{ -1 }$ \\
                    & & z & 0.24 & 0.41 & 0.53 & 0.62 & 0.69 \\
                    & & $p \; [\mathrm{yr}]$ & $1.4 \times10^{ 2 }$ & $1.5 \times10^{ 3 }$ & $9.6 \times10^{ 3 }$ & $4.0 \times10^{ 4 }$ & $2.0 \times10^{ 5 }$ \\
                    & & $a \; [\mathrm{pc}]$ & $2.1 \times10^{ -2 }$ & $1.3 \times10^{ -1 }$ & $4.9 \times10^{ -1 }$ & $1.3 \times10^{ 0 }$ & $4.1 \times10^{ 0 }$ \\
                    & & $a \; [r_g]$ & $1.2 \times10^{ 4 }$ & $1.4 \times10^{ 5 }$ & $4.5 \times10^{ 5 }$ & $1.2 \times10^{ 6 }$ & $4.1 \times10^{ 6 }$ \\
                \hline
                \multirow{4}{1.5cm}{Secondary\newline Offset\newline($\voff$)} & \multirow{4}{*}{$0.49\%$}
                    & $M \; [M_\odot]$ & $1.4 \times10^{ 8 }$ & $4.0 \times10^{ 8 }$ & $7.6 \times10^{ 8 }$ & $1.4 \times10^{ 9 }$ & $3.7 \times10^{ 9 }$ \\
                    & & q & $9.6 \times10^{ -4 }$ & $4.2 \times10^{ -3 }$ & $1.2 \times10^{ -2 }$ & $3.6 \times10^{ -2 }$ & $1.5 \times10^{ -1 }$ \\
                    & & z & 0.25 & 0.42 & 0.54 & 0.63 & 0.69 \\
                    & & $p \; [\mathrm{yr}]$ & $8.3 \times10^{ 2 }$ & $1.5 \times10^{ 3 }$ & $2.3 \times10^{ 3 }$ & $3.7 \times10^{ 3 }$ & $9.0 \times10^{ 3 }$ \\
                    & & $a \; [\mathrm{pc}]$ & 0.22 & 0.41 & 0.58 & 0.87 & 1.82 \\
                    & & $a \; [r_g]$ & $5.0 \times10^{ 3 }$ & $1.1 \times10^{ 4 }$ & $1.8 \times10^{ 4 }$ & $2.9 \times10^{ 4 }$ & $4.9 \times10^{ 4 }$ \\
                \hline
                \multirow{4}{1.5cm}{Secondary\newline Changing\newline($\dvel$)} & \multirow{4}{*}{$0.027\%$}
                    & $M \; [M_\odot]$ & $2.9 \times10^{ 8 }$ & $8.7 \times10^{ 8 }$ & $1.6 \times10^{ 9 }$ & $2.7 \times10^{ 9 }$ & $5.5 \times10^{ 9 }$ \\
                    & & q & $3.9 \times10^{ -4 }$ & $7.3 \times10^{ -4 }$ & $2.2 \times10^{ -3 }$ & $3.8 \times10^{ -3 }$ & $9.1 \times10^{ -3 }$ \\
                    & & z & 0.25 & 0.37 & 0.50 & 0.59 & 0.68 \\
                    & & $p \; [\mathrm{yr}]$ & $5.2 \times10^{ 2 }$ & $7.0 \times10^{ 2 }$ & $8.6 \times10^{ 2 }$ & $1.1 \times10^{ 3 }$ & $1.4 \times10^{ 3 }$ \\
                    & & $a \; [\mathrm{pc}]$ & 0.19 & 0.30 & 0.39 & 0.49 & 0.73 \\
                    & & $a \; [r_g]$ & $2.5 \times10^{ 3 }$ & $3.7 \times10^{ 3 }$ & $5.6 \times10^{ 3 }$ & $8.0 \times10^{ 3 }$ & $1.5 \times10^{ 4 }$ \\
                \hline
                \multirow{4}{1.5cm}{Secondary Both\newline ($\voff$ \& $\dvel$)} & \multirow{4}{*}{$0.014\%$}
                    & $M \; [M_\odot]$ & $4.4 \times10^{ 8 }$ & $1.2 \times10^{ 9 }$ & $1.8 \times10^{ 9 }$ & $3.0 \times10^{ 9 }$ & $6.1 \times10^{ 9 }$ \\
                    & & q & $3.6 \times10^{ -4 }$ & $6.2 \times10^{ -4 }$ & $1.8 \times10^{ -3 }$ & $3.1 \times10^{ -3 }$ & $5.7 \times10^{ -3 }$ \\
                    & & z & 0.25 & 0.37 & 0.48 & 0.58 & 0.68 \\
                    & & $p \; [\mathrm{yr}]$ & $5.2 \times10^{ 2 }$ & $7.0 \times10^{ 2 }$ & $8.4 \times10^{ 2 }$ & $1.0 \times10^{ 3 }$ & $1.4 \times10^{ 3 }$ \\
                    & & $a \; [\mathrm{pc}]$ & 0.22 & 0.33 & 0.41 & 0.51 & 0.76 \\
                    & & $a \; [r_g]$ & $2.3 \times10^{ 3 }$ & $3.3 \times10^{ 3 }$ & $4.8 \times10^{ 3 }$ & $6.6 \times10^{ 3 }$ & $1.1 \times10^{ 4 }$ \\
                \hline
            \end{tabular}
            \captionof{table}{\textbf{Parameters of detectable binary systems with redshift $z < 0.7$.}  The indicated quantiles are given for total mass ($M$), mass ratio ($q$), separation ($a$), and orbital period ($p$).  In only four binaries ($2\E{-8}$ of systems) is the primary detectable, and those have parameters: \mbox{$M\approx 2$--$5\E{9} \, \msol$},
            \hspace{0.5ex} \mbox{$q \approx 0.7$--$0.9$}, \hspace{0.5ex} \mbox{$z \approx 0.6$--$0.8$}, \hspace{0.5ex} \mbox{$a \approx 2.2$--$2.8 \, \pc$}.}
            \label{tab:obs_pars}

        \end{minipage}

    \twocolumn

\end{document}